\documentclass[english,a4paper,final,twocolumn,superscriptaddress]{revtex4-2}
\usepackage[T1]{fontenc}
\usepackage{graphicx}
\usepackage{mathtools}
\usepackage{amssymb}
\usepackage{xcolor}
\usepackage{babel}
\usepackage{subfig}
\usepackage{hyperref}
\usepackage{epstopdf}
\usepackage{subcaption}
\begin{document}
	\title{Thermal transport and low-temperature specific heat in 4Hb-TaS$_2$}

\author{M. Gillig}
\thanks{These authors contributed equally to this work.}
\affiliation{Leibniz Institute for Solid State and
	Materials Research IFW Dresden, Helmholtzstra\ss e 20, 01069 Dresden, Germany}

\author{I.\, Mangel}
\thanks{These authors contributed equally to this work.}
\affiliation{Department of
	Physics, Technion, Haifa, 3200003, Israel }

    \author{I.\, Feldman}
\affiliation{Department of
	Physics, Technion, Haifa, 3200003, Israel }

\author{V. Kocsis}
\affiliation{Leibniz Institute for Solid State and
	Materials Research IFW Dresden, Helmholtzstra\ss e 20, 01069 Dresden, Germany}

\author{A.\,Kanigel}
\affiliation{Department of
	Physics, Technion, Haifa, 3200003, Israel }

%
%

	\author{D.\,V.~Efremov}

\affiliation{Leibniz Institute for Solid State and
	Materials Research IFW Dresden, Helmholtzstra\ss e 20, 01069 Dresden, Germany}

\author{B. B\"uchner}
\affiliation{Leibniz Institute for Solid State and
	Materials Research IFW Dresden, Helmholtzstra\ss e 20, 01069 Dresden, Germany}
\affiliation{Würzburg-Dresden Cluster of Excellence ct.qmat,
	Technische Universität Dresden, 01069 Dresden, Germany}
	
	\begin{abstract} 
		
		We investigate the low-energy excitation spectrum of the van der Waals heterostructure superconductor 4Hb-TaS$_2$ using ultra-low-temperature specific-heat and thermal-conductivity measurements with magnetic fields applied parallel and perpendicular to the crystallographic $c$ axis. The specific heat is broadly consistent with a nodeless superconducting gap, but retains a finite residual linear contribution, indicating a small residual low-energy density of states in the superconducting state. In addition, a pronounced upturn appears below approximately 0.3~K. Its weak magnetic-field dependence, together with the absence of a corresponding feature in thermal transport, supports an interpretation in terms of localized degrees of freedom, most likely a nuclear Schottky contribution. In contrast to the finite residual thermodynamic density of states, the thermal conductivity extrapolates to a vanishing zero-field electronic linear term within experimental uncertainty for both field orientations. Thus, the residual low-energy states do not form a detectable itinerant heat-conduction channel. In finite magnetic field, the electronic heat transport grows rapidly. For out-of-plane fields, this response is broadly consistent with previous thermal-conductivity measurements and with the behavior commonly associated with multigap nodeless superconductivity. The even steeper increase observed for in-plane fields suggests that the field-induced quasiparticle response of 4Hb-TaS$_2$ is more complicated than the standard multigap picture alone.

	\end{abstract}
	
	\maketitle
	\section{Introduction}

 Significant research efforts are currently focused on discovering novel superconductors with topologically non-trivial order parameters, motivated by their potential for fault-tolerant quantum information processing  \cite{Qi2011,Alicea2012,Sato2017,Nayak2008,Li2019}. Several candidate systems have been proposed, including Sr\textsubscript{2}RuO\textsubscript{4}, UPt\textsubscript{3}, UTe\textsubscript{2}, and SrPtAs \cite{Kallin2012,Kallin2016,Mackenzie2017,Grinenko2021,Jiao2020,Machida2021,Mandal2023}. However,  the nature of their superconducting order parameters remains controversial. This uncertainty makes exploring new candidate materials in condensed matter physics critical.

In a recent study,  4Hb-TaS$_2$ was identified as a promising candidate for hosting chiral topological superconductivity  \cite{Ribak2020}.  This compound belongs to the TaS$_2$ family, which includes several polymorphs. The best known are the superconducting semimetal 2H-TaS$_2$ with a superconducting critical temperature of 0.8 K and the strongly correlated Mott insulator with a spin-liquid ground state 1T-TaS$_2$  \cite{Fazekas1980,Law2017,Ribak2017,Murayama2018}. 
 4Hb-TaS$_2$ may be viewed as a natural layered heterostructure consisting of
 weakly coupled quasi-two-dimensional 1T-TaS$_2$ and 1H-TaS$_2$ monolayers. 
  Because the interlayer coupling is relatively weak, the ARPES spectra can be understood approximately
 in terms of the superposed band structures of the two constituent layers.

Superconductivity in 4Hb-TaS$_2$ was reported by Ribak \textit{et al.}, with a
transition temperature of $T_c \approx 2.7$~K \cite{Ribak2020}. Moreover, their zero-field $\mu$SR measurements showed the emergence of spontaneous internal fields below $T_c$, providing evidence for time-reversal-symmetry breaking in the superconducting state \cite{Ribak2020}. The same study also demonstrated pronounced anisotropy and
quasi-two-dimensional behavior, indicating 4Hb-TaS$_2$ as a promising
candidate for realizing exotic superconducting phases.

The superconducting ground state of 4Hb-TaS$_2$ remains unsettled despite a rapidly
expanding body of experimental work. Broken time-reversal symmetry was inferred from
zero-field $\mu$SR measurements, while scanning tunneling spectroscopy revealed
topological boundary modes and finite in-gap spectral weight. Subsequent experiments
reported magnetic memory and spontaneous vortices, two-component nematic
superconductivity, and $\pi$-shifted Little--Parks oscillations, all of which point to
an unconventional and possibly multicomponent order parameter
\cite{Ribak2020,Nayak2021,Persky2022,Silber2024,Almoalem2024}. In parallel, ARPES and
theory have emphasized strong layer selectivity, charge transfer, reduced interlayer
hopping, and interlayer-pairing scenarios that can naturally generate strongly
band-dependent superconducting responses \cite{Almoalem2024,Dentelski2021,Liu2024}.

Recent experimental studies have raised an important question regarding the low-energy excitation spectrum of 4Hb-TaS$_2$. Penetration-depth measurements have been interpreted in terms of a fully gapped, nodeless two-band superconducting state \cite{Zhou2025}. In contrast, ultra-low-temperature thermal-conductivity measurements reported a finite residual linear term, corresponding to approximately 2.5\% of the normal-state value, and interpreted it as evidence for a small population of mobile gapless quasiparticles surviving deep in the superconducting state \cite{Wang2025}. At first sight, these observations are difficult to reconcile: can a superconductor have fully open gaps on all Fermi-surface sheets while simultaneously supporting a finite density of mobile zero-energy quasiparticles?

To address this question, we performed an independent study of both low-temperature specific heat and thermal conductivity on high-quality 4Hb-TaS$_2$ crystals from the same growth batch used in our previous studies. This combination allows us to compare the thermodynamic low-energy density of states with the itinerant quasiparticle contribution to heat transport. As shown below, our thermal-conductivity data extrapolate to a vanishing residual linear term within experimental uncertainty, placing strong constraints on any residual metallic component in the superconducting state.

\section{Experimental Methods} 

\subsection{Samples}

High-quality single crystals of 4Hb-TaS$_2$ were grown by the chemical vapor transport (CVT) method. To improve the crystal quality and increase the superconducting volume fraction, 1\% of the sulfur was substituted with selenium, following the procedure reported in Ref.~\cite{Ribak2020}. The crystal stoichiometry was verified by energy-dispersive x-ray spectroscopy (EDS), while the crystal structure was confirmed by x-ray diffraction (XRD) and transmission electron microscopy (TEM).  These crystals have been extensively characterized previously by $\mu$SR, STM, ARPES, transport and scanning SQUID measurements, demonstrating high crystalline quality and reproducible superconducting properties \cite{Ribak2020,Nayak2021,Persky2022,Almoalem2024,Almoalem2024a}.

\subsection{Specific heat}
 The magnetic-field dependence of the specific heat was measured on two
4Hb-TaS$_2$ samples from the same batch using the PPMS dilution refrigerator insert. 
For out-of-plane field measurements, a $10$~mg sample was measured between $0.1$ and $4$~K. For
in-plane field measurements, a larger 4Hb-TaS$_2$ crystal was wire-cut parallel
to the $c$ axis to obtain a $16.4$~mg sample with a flat surface parallel to
the $c$ axis. The in-plane
field measurements were carried out between $0.15$ and $4$~K.

\subsection{Thermal conductivity}
 Heat transport measurements were conducted in a dilution refrigerator setup (Kelvinox MX-400, Oxford Instruments) equipped with a 16\,T superconducting magnet.
Thermal conductivity was measured with the heat current applied parallel to the ab plane. Magnetic fields were applied either parallel or perpendicular to the crystallographic c axis.
The measurement used the standard four-wire configuration with a heater chip and two thermometers. 
For the high-sensitivity readout of the in-situ calibrated RuO$_2$ chip thermometers, we used an AC resistance bridge (Model 372, Lakeshore).
The temperature-dependent thermal conductivity was measured using a steady-state method, in which signals with the heater on and off were recorded, and the thermal conductivity was calculated from the difference.

\section{results}
\subsection{Specific heat}

Fig.~1 summarizes the temperature dependence of the specific heat of 4Hb-TaS$_2$ for magnetic fields applied parallel (Fig.~1(a)) and  perpendicular  (Fig.~1(b)) to the crystallographic $c$ axis. Fig.~1(c) shows the heat capacity after subtracting the phonon contribution. The data are generally consistent with a nodeless superconducting gap, but retain a finite residual linear contribution at the lowest temperatures, corresponding to a residual Sommerfeld coefficient of $\gamma_r/\gamma \approx 0.1$. Thus, while the majority of the electronic density of states becomes gapped below $T_c$, a small residual low-energy density of states persists in the superconducting state.

After subtracting the residual linear contribution, the electronic specific heat is well described by a single-gap BCS expression with a zero-temperature gap of $\Delta(0)=0.45$~meV, corresponding to $2\Delta(0)/k_BT_c \approx 3.8$, close to the weak-coupling BCS value of 3.52. The temperature dependence of the superconducting electronic specific heat is therefore consistent with a fully gapped, nodeless superconducting state. The extracted gap value is in good agreement with previous reports on 4Hb-TaS$_2$~\cite{Ribak2020}.

To analyze the low-temperature electronic contribution, we subtract the phonon contribution, $C_{\rm ph}=\beta T^3$, from the total specific heat and define
$$
C_e \equiv C-C_{\rm ph}= \gamma T + C_{\rm Schottky}.
$$
Here, the first term represents the electronic contribution, while $C_{\rm Schottky}$ accounts for the low-temperature Schottky anomaly. The resulting phonon-subtracted heat capacity is shown in Fig. 1(c). Fitting the normal-state data above $H_{c2}$ yields $\gamma=5.7$~mJ,mol$^{-1}$,K$^{-2}$ and $\beta=0.388$~mJ,mol$^{-1}$,K$^{-4}$. The detailed fitting procedure and the analysis of the individual contributions are presented in the Supplementary Material.

The electronic heat capacity exhibits a pronounced upturn below approximately 0.3~K. In contrast to the corresponding anomaly reported in 1T-TaS$_2$~\cite{Ribak2017}, which displays a strong magnetic-field dependence consistent with an electronic origin, the low-temperature anomaly in 4Hb-TaS$_2$ evolves only weakly with magnetic field. Although a nuclear Schottky anomaly is generally expected to exhibit a Zeeman-induced field dependence, the $^{181}$Ta nucleus ($I=7/2$) possesses a large nuclear quadrupole moment, resulting in a substantial quadrupolar splitting that dominates over the Zeeman energy in the magnetic-field range explored here. A calculation based on the measured NQR parameters reproduces the observed temperature dependence of the anomaly (see Supplementary Material), supporting its assignment to a nuclear Schottky contribution.

\begin{figure}[t!]
	\centering	\subfloat{\textbf{a)}}{\includegraphics[width = 0.95\linewidth]{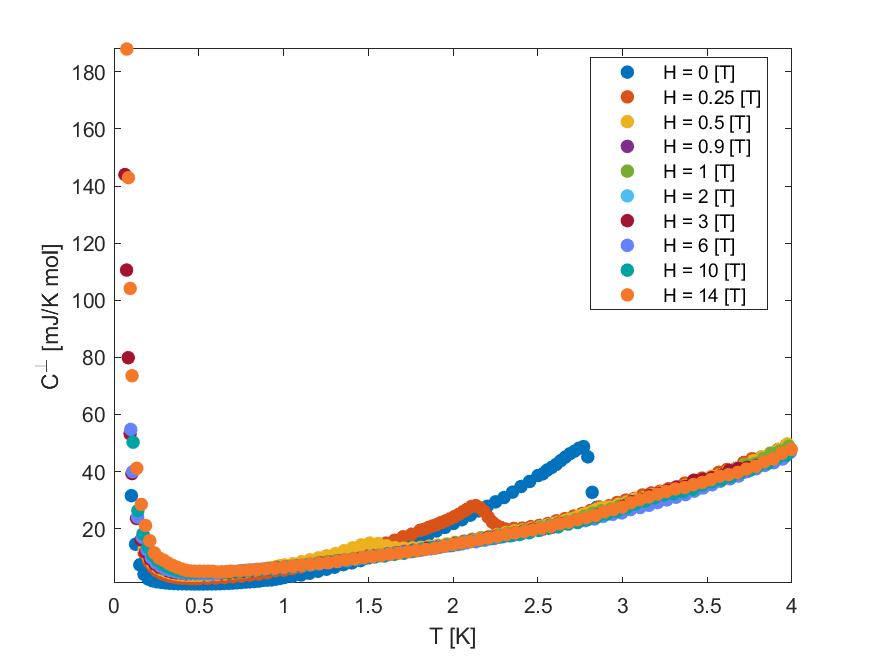}}\\
    \subfloat{\textbf{b)}}{\includegraphics[width=0.95\linewidth]{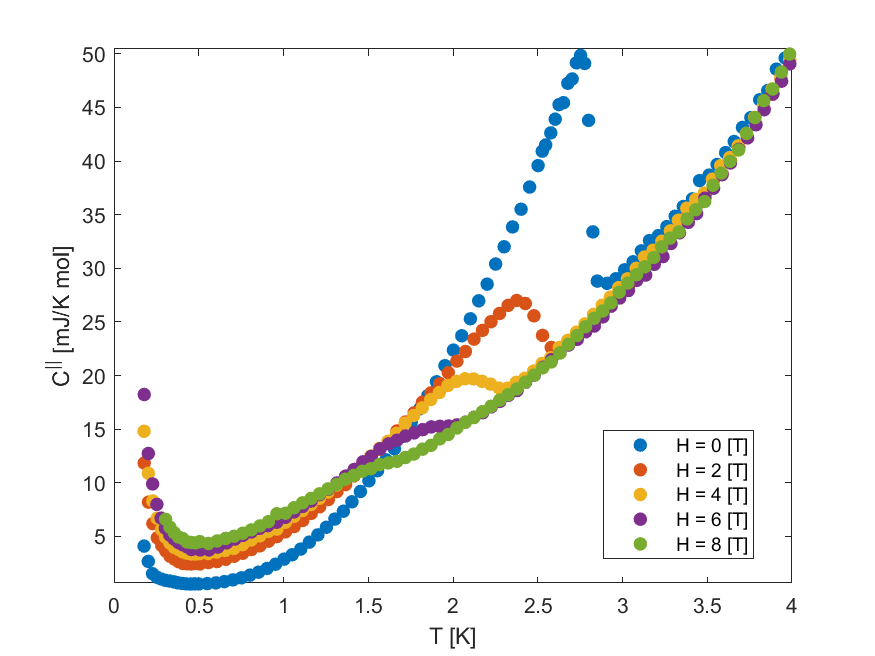}} \\ \subfloat{\textbf{c)}}{\includegraphics[width=0.95\linewidth]{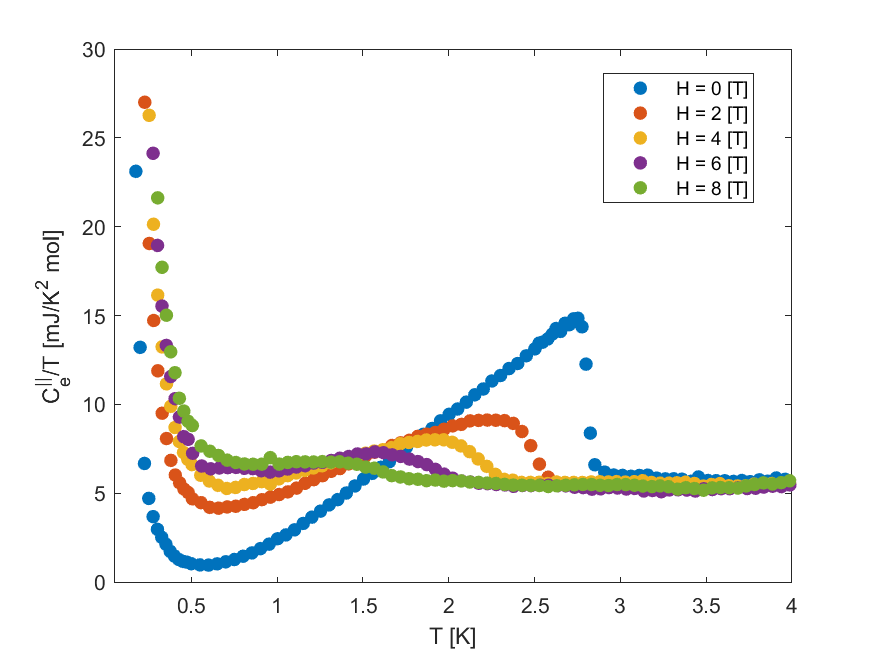}}
	\caption[Specific heat]{ Temperature dependence of the specific heat of 4Hb-TaS$_2$ measured in magnetic fields applied 
(a) out of plane, $H\parallel c$, and 
(b) in plane, $H\perp c$. 
(c) Specific heat after subtracting the phonon contribution, $C-C_{\rm ph}$. }
	\label{fi:gamma_beta_fits}
\end{figure}

\begin{figure*}[t!]
	\centering
	\includegraphics[width=0.95\linewidth]{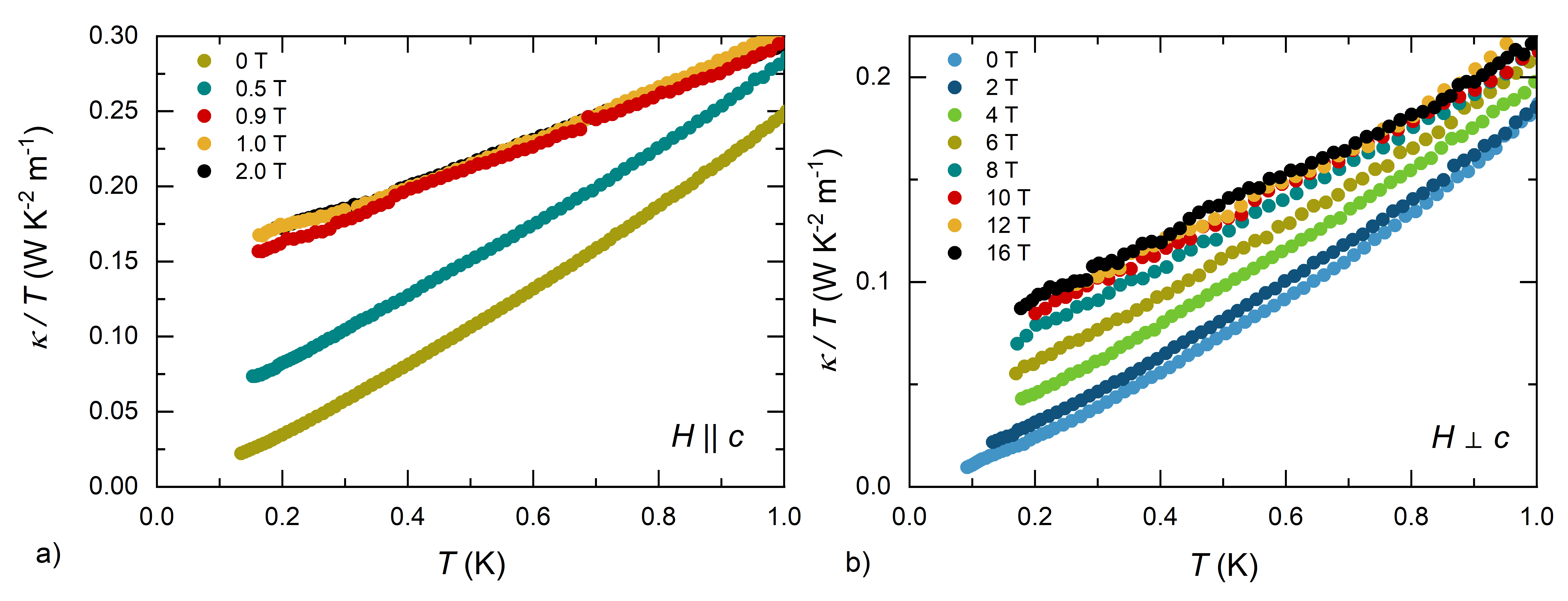}
	\caption{ Temperature dependence of $\kappa/T$ for magnetic fields applied 
(a) parallel and 
(b) perpendicular to the crystallographic $c$ axis. 
The magnetic field strengths are indicated in the legend.}
	\label{fig:figkappatvsh}
\end{figure*}

\subsection{Low-temperature thermal conductivity}

Fig.~\ref{fig:figkappatvsh} shows $\kappa/T$ as a function of temperature for magnetic fields applied parallel and perpendicular to the crystallographic $c$ axis. In zero magnetic field, the data extrapolate to a vanishing residual linear term within the experimental uncertainty. This indicates that, despite the finite residual density of states inferred from the specific heat, there is no detectable residual metallic heat-conduction channel in the superconducting ground state. For fields above the corresponding upper critical fields, $H_{c2}^{\parallel c}\simeq 1$~T and $H_{c2}^{\perp c}\simeq 14$--$16$~T, in agreement with the values reported in Ref.~\cite{Ribak2020}, the curves become nearly field independent, consistent with the recovery of the normal-state thermal conductivity.

To quantify the residual electronic heat transport, we analyze the low-temperature thermal conductivity using
\begin{equation}
\kappa(T,H)=\alpha(H)T+\kappa_{\rm ph}(T,H),
\label{eq:kappa_decomp}
\end{equation}
where $\alpha(H)T$ is the electronic contribution and $\kappa_{\rm ph}$ is the phonon contribution. Since the phonon term is sizable and does not follow a simple $T^3$ law over the measured temperature range, an accurate effective description of $\kappa_{\rm ph}$ is essential for extracting a reliable value of $\alpha(H)$.

 Since the specific-heat analysis yields a conventional phonon term, $C_{\rm ph}=\beta T^3$, the sub-cubic temperature dependence of the phonon thermal conductivity must originate from the phonon mean free path rather than from an anomalous phonon density of states. Within the kinetic expression
\[
\kappa_{\rm ph}=\frac{1}{3}C_{\rm ph}\langle v_{\rm ph}\rangle l_{\rm ph},
\]
the observed effective power law therefore implies a temperature-dependent $l_{\rm ph}$. This motivates the phenomenological form of $\kappa_{\rm ph}$ introduced below.

We describe the phonon contribution phenomenologically by allowing the phonon mean free path to be limited by boundary scattering and an additional field-dependent scattering channel, combined through Matthiessen's rule:
\begin{equation}
\kappa_{\rm ph}(T,H)=
\left(
\frac{1}{\eta T^\lambda}
+
\frac{b(H)}{T^2}
\right)^{-1}.
\end{equation}
Here, $\eta$ and $\lambda$ describe the zero-field phonon background, while $b(H)$ captures the field-induced modification of the phonon mean free path. This model is used to separate the electronic linear term from the phonon background and should be viewed as a phenomenological transport decomposition rather than a microscopic model of a unique phonon-scattering process.

\begin{figure}
	\centering	
	\subfloat{\textbf{a)}}{\includegraphics[width=0.8\linewidth]{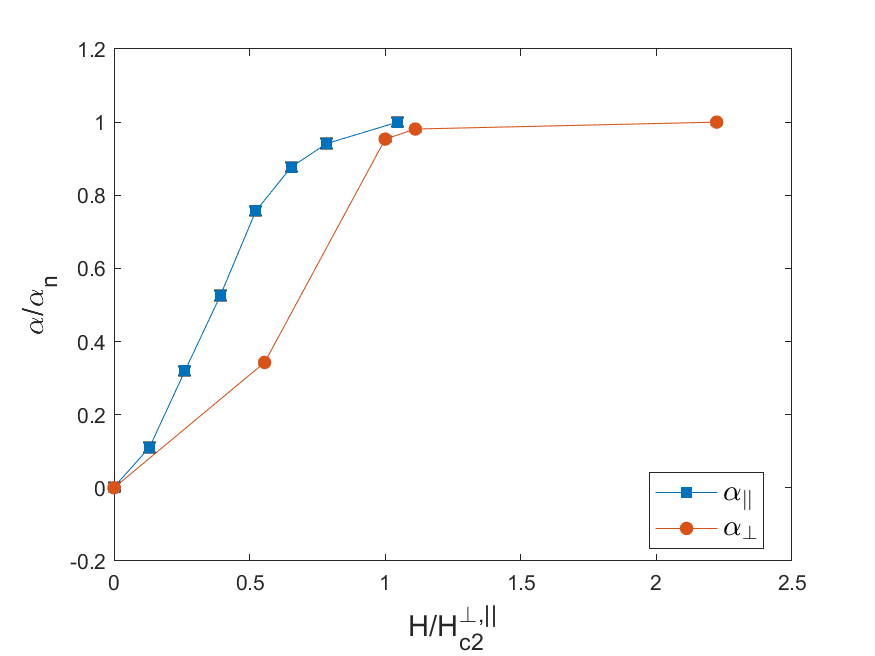}}
	\subfloat{\textbf{b)}}{\includegraphics[width=0.8\linewidth]{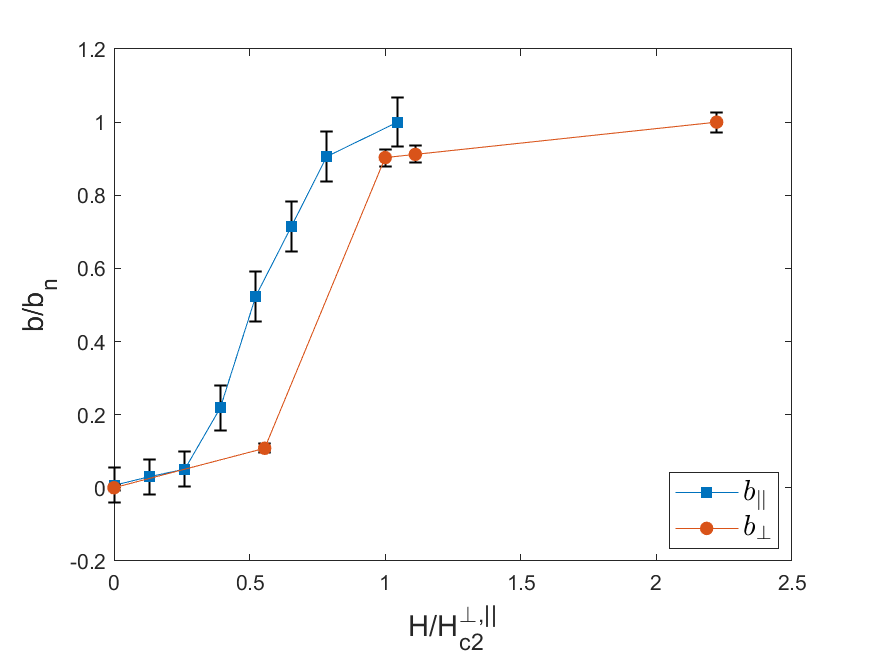}}
	\caption{ Magnetic-field dependence of the fitting parameters obtained from the phenomenological thermal-conductivity analysis. 
(a) Electronic heat-transport coefficient $\alpha(H)$, normalized by its high-field normal-state value $\alpha_N$. 
(b) Field-dependent phonon-scattering parameter $b(H)$, normalized by its high-field value. 
The data are plotted as a function of $H/H_{c2}$ for fields applied parallel and perpendicular to the crystallographic $c$ axis.}
	\label{fig:fig3}
\end{figure}

The zero-field data were first fitted with $b(0)=0$, fixing the phonon-background parameters $\eta$ and $\lambda$. These parameters were then kept fixed for finite-field fits, while $\alpha(H)$ and $b(H)$ were allowed to vary. The resulting $\alpha(H)$ is shown in Fig.~\ref{fig:fig3}(a). In zero magnetic field, $\alpha(0)$ is zero within the experimental uncertainty for both field orientations, placing an upper bound on any residual metallic heat-conduction channel that is well below the 2.5\% normal-state value reported in Ref.~\cite{Wang2025}. Thus, although the specific heat reveals a finite residual density of states, these states do not produce a detectable itinerant heat-current response.

The field-dependent parameter $b(H)$, shown in Fig.~\ref{fig:fig3}(b), describes the additional field-induced contribution to the phonon thermal resistance in our phenomenological model. It is therefore essential for obtaining a reliable separation between the phonon background and the electronic linear term. In both field orientations, $b(H)$ is negligible in zero field and grows as the superconducting state is suppressed by magnetic field, indicating that the phonon mean free path is modified by field-induced low-energy excitations. However, we do not interpret $b(H)$ uniquely as scattering from conventional vortex cores. In a layered superconductor, the vortex structure is expected to be very different for fields applied parallel and perpendicular to the crystallographic $c$ axis. We therefore regard $b(H)$ as a phenomenological parameter that captures the field-induced modification of the phonon background, while the electronic coefficient $\alpha(H)$ provides the primary measure of itinerant quasiparticle heat transport.

\begin{figure}
	\centering
	\includegraphics[width=1.8\columnwidth]{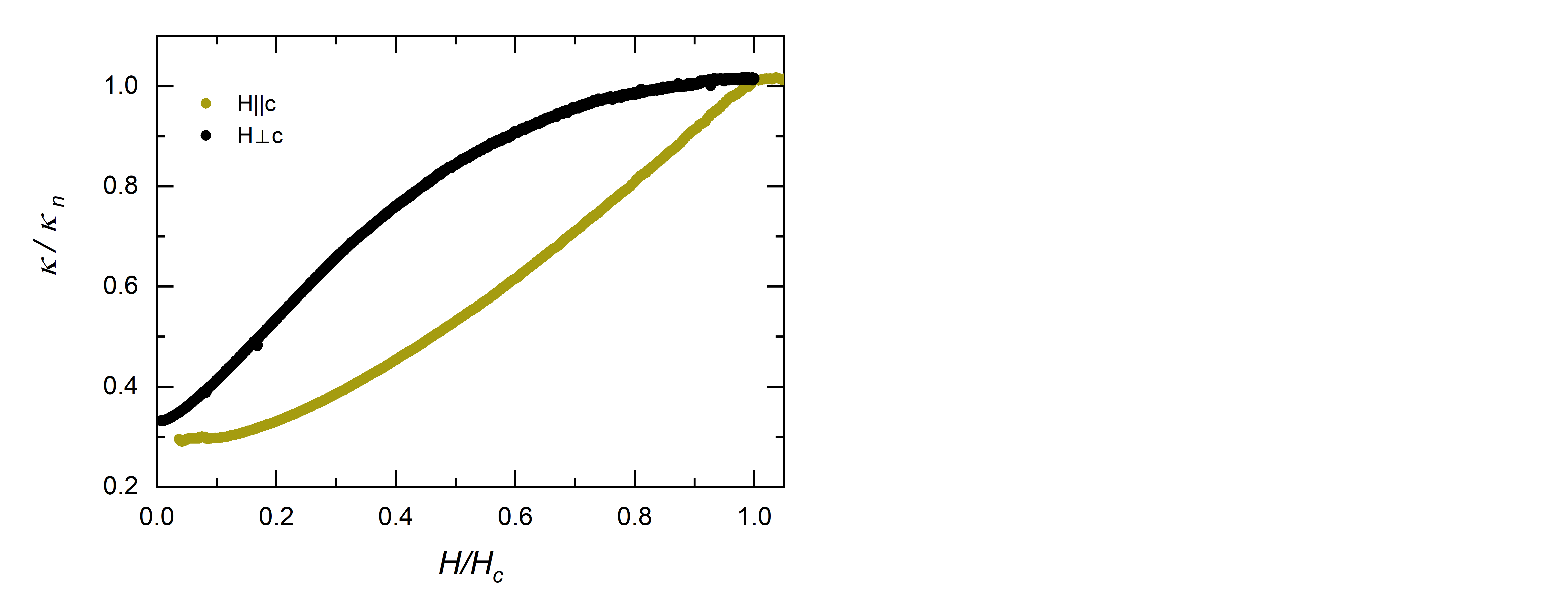}
	\caption{Magnetic-field dependence of the normalized thermal conductivity $\kappa/\kappa_n$
		at T=300 mK for out-of-plane field $ H||c$ and in-plane field  $H\perp c$.}
	\label{fig:fig5a}
\end{figure}

Fig. ~\ref{fig:fig5a} shows the magnetic-field dependence of the normalized thermal conductivity, $\kappa/\kappa_n$, at $T\simeq 300$~mK. This representation gives the same qualitative field dependence as the extracted electronic coefficient $\alpha(H)$ shown in Fig.~\ref{fig:fig3}(a). For out-of-plane fields, $H\parallel c$, both quantities show a relatively rapid increase with field, indicating the growth of a field-induced itinerant quasiparticle contribution. Such a rapid recovery of electronic heat transport at fields well below $H_{c2}$ is commonly taken as a signature of multigap nodeless superconductivity, where quasiparticles associated with a smaller superconducting gap are activated at relatively low fields. In this respect, our out-of-plane-field data are broadly consistent with the results of Wang \textit{et al.}~\cite{Wang2025}. The more surprising observation is that the increase is even steeper for in-plane fields. This pronounced field-orientation dependence suggests that the finite-field quasiparticle response of 4Hb-TaS$_2$ is more complicated than the standard multigap picture alone, and may also involve the layered electronic structure, anisotropic vortex configurations, Zeeman effects, or spin-orbit coupling.

\section{Conclusions}

In conclusion, we have investigated the low-energy excitation spectrum of 4Hb-TaS$_2$ using ultra-low-temperature specific heat and thermal conductivity with magnetic fields applied both parallel and perpendicular to the crystallographic $c$ axis. The specific-heat data are broadly consistent with a nodeless superconducting gap, but retain a finite residual linear contribution, indicating that a small low-energy density of states persists in the superconducting state. In addition, the specific heat exhibits a pronounced upturn below approximately 0.3~K. Its weak magnetic-field dependence, together with the absence of a corresponding anomaly in thermal transport, supports an interpretation in terms of localized degrees of freedom, most likely a nuclear Schottky contribution.

The thermal-conductivity measurements lead to a central distinction between thermodynamic and transport signatures of low-energy excitations. After separating the electronic linear term from the phonon background, the zero-field electronic coefficient $\alpha(0)$ vanishes within experimental uncertainty for both field orientations. Thus, although specific heat reveals a finite residual density of states, these states do not form a detectable itinerant heat-conduction channel. This places a strong constraint on any residual metallic component in the superconducting ground state of 4Hb-TaS$_2$.

In finite magnetic field, the electronic heat transport grows rapidly as superconductivity is suppressed. For out-of-plane fields, the field dependence of $\alpha(H)$ and of the normalized thermal conductivity is broadly consistent with previous thermal-conductivity measurements and resembles the behavior commonly associated with multigap nodeless superconductivity. The even steeper increase observed for in-plane fields, however, suggests that the field-induced quasiparticle response is more complicated than the standard multigap picture alone. This anisotropic response may reflect the layered electronic structure, anisotropic vortex configurations, Zeeman effects, or spin-orbit coupling. Taken together, our results show that 4Hb-TaS$_2$ combines a nodeless superconducting gap structure with residual thermodynamic low-energy spectral weight that is not accompanied by mobile zero-field quasiparticles.

\section{Acknowledgments} 
We acknowledge useful discussions with Jonathan Ruhman. This work was supported by the German–Israeli
Project Cooperation (DIP), Grant Nos. 529677299 and
3970/1-1. 
	
\bibliographystyle{apsrev4-1_title}

\bibliography{4HbTaS2}

\end{document}